\documentclass[twocolumn,twoside,slac_two]{revtex4}
\usepackage{graphicx}
\usepackage{fancyhdr}
\begin{document}

\title{Radio Searches of Fermi LAT Sources and Blind Search Pulsars: \\ The Fermi Pulsar Search Consortium}

%

\author{P.~S.~Ray}
\email{paul.ray@nrl.navy.mil}
\affiliation{Space Science Division, Naval Research Laboratory, Washington, DC 20375-5352, USA}
\author{A.~A.~Abdo}
\affiliation{George Mason University, Fairfax, VA, resident at Naval Research Laboratory, Washington, DC, USA}
\author{D.~Bhattacharya}
\affiliation{Inter-University Centre for Astronomy and Astrophysics, Pune 411 007, India}
\author{B.~Bhattacharyya}
\affiliation{Inter-University Centre for Astronomy and Astrophysics, Pune 411 007, India}
\author{F.~Camilo}
\affiliation{Columbia Astrophysics Laboratory, Columbia University, New York, NY 10027, USA}
\author{I.~Cognard}
\affiliation{LP2CE, CNRS, and Orl\'eans University, F-45071 Orl\'eans Cedex 02,  and Station de radioastronomie de Nan\c{c}ay, Observatoire de Paris et CNRS/INSU, F-18330 Nan\c{c}ay, France}
\author{E.~C.~Ferrara}
\affiliation{NASA Goddard Space Flight Center, Greenbelt, MD 20771, USA}
\author{P.~C.~C.~Freire}
\affiliation{Max-Planck-Institut f\"ur Radioastronomie, Auf dem H\"ugel 69, 53121 Bonn, Germany}
\author{L.~Guillemot}
\affiliation{Max-Planck-Institut f\"ur Radioastronomie, Auf dem H\"ugel 69, 53121 Bonn, Germany}
\author{Y.~Gupta}
\affiliation{National Centre for Radio Astrophysics, TIFR, Pune 411 007, India}
\author{A.~K.~Harding}
\affiliation{NASA Goddard Space Flight Center, Greenbelt, MD 20771, USA}
\author{J.~W.~T.~Hessels}
\affiliation{ASTRON, the Netherlands Institute for Radio Astronomy, Postbus 2, 7990 AA, Dwingeloo, The Netherlands}
\author{S.~Johnston}
\affiliation{CSIRO, Australia Telescope National Facility, Epping NSW 1710, Australia}
\author{M.~Keith}
\affiliation{CSIRO, Australia Telescope National Facility, Epping NSW 1710, Australia}
\author{M.~Kerr}
\affiliation{Stanford University, Stanford, CA, USA}
\author{M.~Kramer}
\affiliation{Jodrell Bank Centre for Astrophysics, University of Manchester, UK}
\affiliation{Max-Planck-Institut f\"ur Radioastronomie, Auf dem H\"ugel 69, 53121 Bonn, Germany}
\author{M.~A.~McLaughlin}
\affiliation{Department of Physics, West Virginia University, Morgantown, WV 26506, USA}
\author{P.~F.~Michelson}
\affiliation{Stanford University, Stanford, CA, USA}
\author{D. Parent}
\affiliation{George Mason University, Fairfax, VA, resident at Naval Research Laboratory, Washington, DC, USA}
\author{S.~M.~Ransom}
\affiliation{National Radio Astronomy Observatory (NRAO), Charlottesville, VA 22903, USA}
\author{M.~S.~E.~Roberts}
\affiliation{Eureka Scientific, Oakland, CA 94602, USA}
\author{R.~W.~Romani}
\affiliation{Stanford University, Stanford, CA, USA}
\author{J.~Roy}
\affiliation{National Centre for Radio Astrophysics, TIFR, Pune 411 007, India}
\author{P.~M.~Saz~Parkinson}
\affiliation{University of California at Santa Cruz, Santa Cruz, CA 95064, USA}
\author{R.~Shannon}
\affiliation{CSIRO, Australia Telescope National Facility, Epping NSW 1710, Australia}
\author{D.~A.~Smith}
\affiliation{Universit\'e Bordeaux 1, CNRS/IN2p3, Centre d'\'Etudes Nucl\'eaires de Bordeaux Gradignan, France}
\author{B.~W.~Stappers}
\affiliation{Jodrell Bank Centre for Astrophysics,The University of Manchester, UK}
\author{G.~Theureau}
\affiliation{LP2CE, CNRS, and Orl\'eans University, F-45071 Orl\'eans Cedex 02,  and Station de radioastronomie de Nan\c{c}ay, Observatoire de Paris et CNRS/INSU, F-18330 Nan\c{c}ay, France}
\author{D.~J.~Thompson}
\affiliation{NASA Goddard Space Flight Center, Greenbelt, MD 20771, USA}
\author{P.~Weltevrede}
\affiliation{Jodrell Bank Centre for Astrophysics,The University of Manchester, UK}
\author{K.~S.~Wood}
\affiliation{Space Science Division, Naval Research Laboratory, Washington, DC 20375, USA}
\author{M.~Ziegler}
\affiliation{University of California at Santa Cruz, Santa Cruz, CA 95064, USA}

\begin{abstract}
We present a summary of the \textit{Fermi} Pulsar Search Consortium (PSC), an international collaboration of radio astronomers and members of the Large Area Telescope (LAT) collaboration, whose goal is to organize radio follow-up observations of \textit{Fermi} pulsars and pulsar candidates among the LAT $\gamma$-ray source population. The PSC includes pulsar observers with expertise using the world's largest radio telescopes that together cover the full sky. We have performed very deep observations of all 35 pulsars discovered in blind frequency searches of the LAT data, resulting in the discovery of radio pulsations from four of them. We have also searched over 300 LAT $\gamma$-ray sources that do not have strong associations with known $\gamma$-ray emitting source classes and have pulsar-like spectra and variability characteristics. These searches have led to the discovery of a total of 43 new radio millisecond pulsars (MSPs) and four normal pulsars. These discoveries greatly increase the known population of MSPs in the Galactic disk, more than double the known population of so-called `black widow' pulsars, and contain many promising candidates for inclusion in pulsar timing arrays.
\end{abstract}

\maketitle

\thispagestyle{fancy}


\section{The \textit{Fermi} Pulsar Search Consortium}

The \textit{Fermi} Pulsar Search Consortium (PSC) was established as a collaborative effort between the 
\textit{Fermi} LAT team and an international group of radio pulsar observers with expertise using radio 
telescopes around the world.  The two goals of the PSC are: (1) to obtain radio observations of $\gamma$-ray pulsars discovered in blind searches of LAT data, (2) to search for radio pulsars in the directions of unassociated LAT sources that have pulsar-like characteristics.  This collaboration has proven extremely successful, as we describe below.  Members of the PSC have used the following radio telescopes for these observations: Green Bank Telescope (GBT), Parkes, Effelsberg, Nan\c{c}ay Radio Telescope (NRT), Arecibo Telescope, the Lovell Telescope at Jodrell Bank, and the Giant Metrewave Radio Telescope (GMRT).

\section{Radio Searches of Blind Search Pulsars}

One of the great successes of the \textit{Fermi} LAT has been the discovery of 35 pulsars in 
blind searches of the $\gamma$-ray data alone \cite{a+09,sdz+10,pga+12,sp+12}.  Although these 
discoveries were made using the $\gamma$-ray data alone, before these new pulsars can be dubbed 
``radio quiet'', they must first be searched deeply for radio pulsations. In addition, any radio 
detections provide important new information on each pulsar including a distance determination 
from the dispersion measure (DM), which is critical for  converting measured $\gamma$-ray fluxes 
into luminosities and thus determining the $\gamma$-ray efficiency of the pulsar. A radio pulse 
detection also yields information on the location of the emission region from the radio to 
$\gamma$-ray offset \cite{rw10}. Radio polarization studies also provide potentially important 
information on the pulsar geometry (e.g. the angles between the rotation and magnetic axes [$
\alpha$] and between the rotation axis and the line-of-sight [$\zeta$]) \cite{rc69}. Radio limits and detections also provide critical input 
for population models, where the blind search pulsars give a much more complete sample of the 
young pulsar population, owing to the broad $\gamma$-ray beams \cite{wr11,twc11}.

\begin{figure}
\includegraphics[width=3.5in]{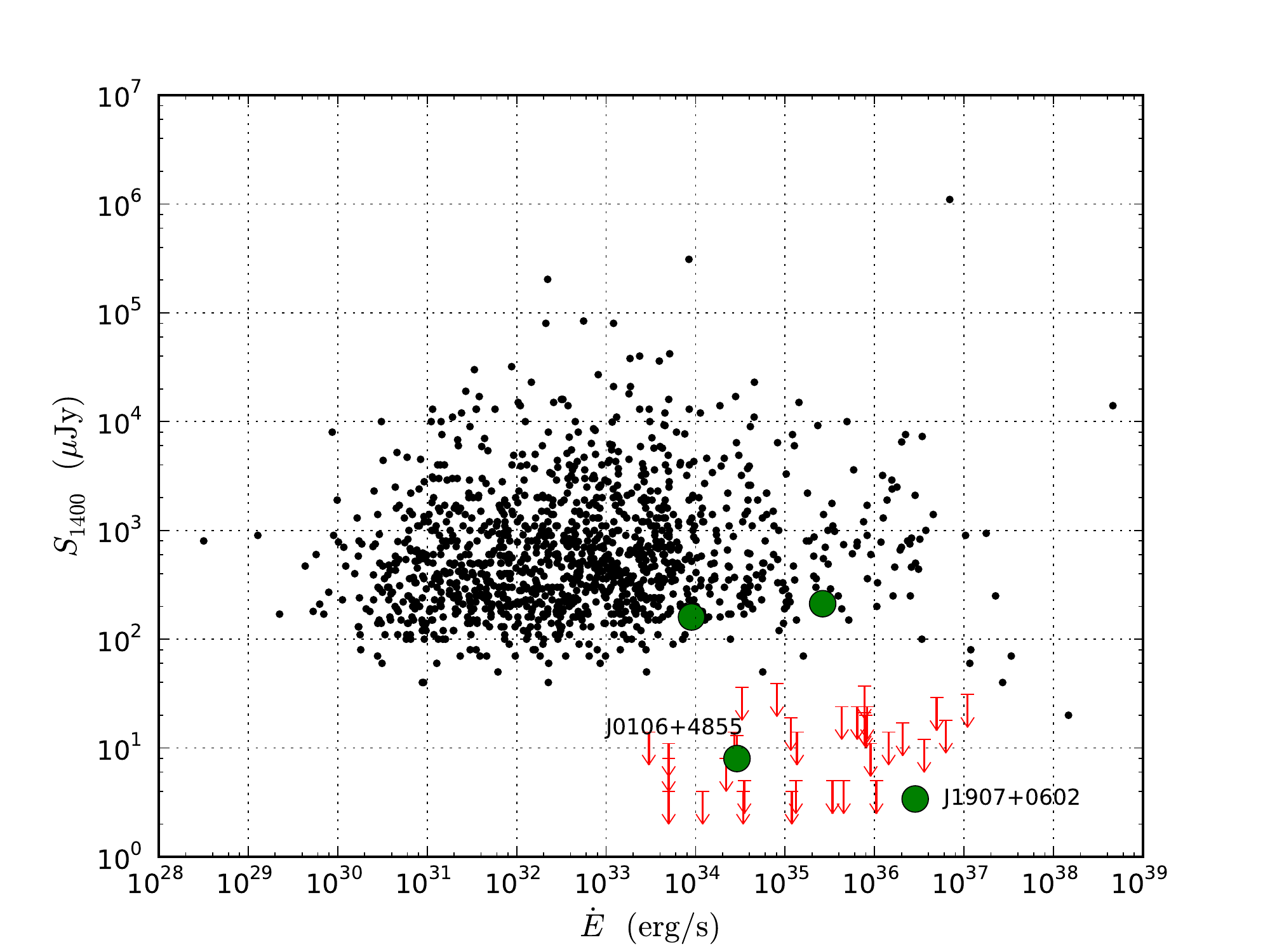}
\caption{Radio flux density upper limits (red arrows) and detections (green circles) for the 35 LAT blind search pulsars, scaled to 1400 MHz using a spectral index of $-1.6$.  The 1400 MHz flux densities of young pulsars in the ATNF catalog are plotted as black dots, for comparison.\label{fig:flux}}
\end{figure}

The PSC has made extremely deep radio observations of all of the LAT-discovered blind search pulsars and 
detected radio pulsations from four of them \cite{crr+09,a+10,pga+12}.
Now, with 4 detections and 31 upper limits, it is appropriate to take a look at  our progress so far. In 
Figure \ref{fig:flux}, we show the flux limits and detections (scaled to 1400 MHz using a typical pulsar 
spectral index of $-1.6$) from our observations. It is clear that we are achieving flux limits competitive 
with the faintest known radio pulsars (all well under 0.1 mJy) and two of our detections (PSR J1907+0602 
and J0106+4855) are very faint indeed!   We also note that a possible (non-PSC) radio detection of PSR J1732$-$3131 has been reported at 34.5 MHz \cite{mad12} and so additional searches at low frequency are certainly warranted.

The vast majority of these pulsars are so faint in the radio that without \textit{Fermi} they would have remained undetected for decades or longer.  In fact, several of them resolved the mystery of what was powering unidentified TeV sources.

\begin{figure}
\includegraphics[width=3.5in]{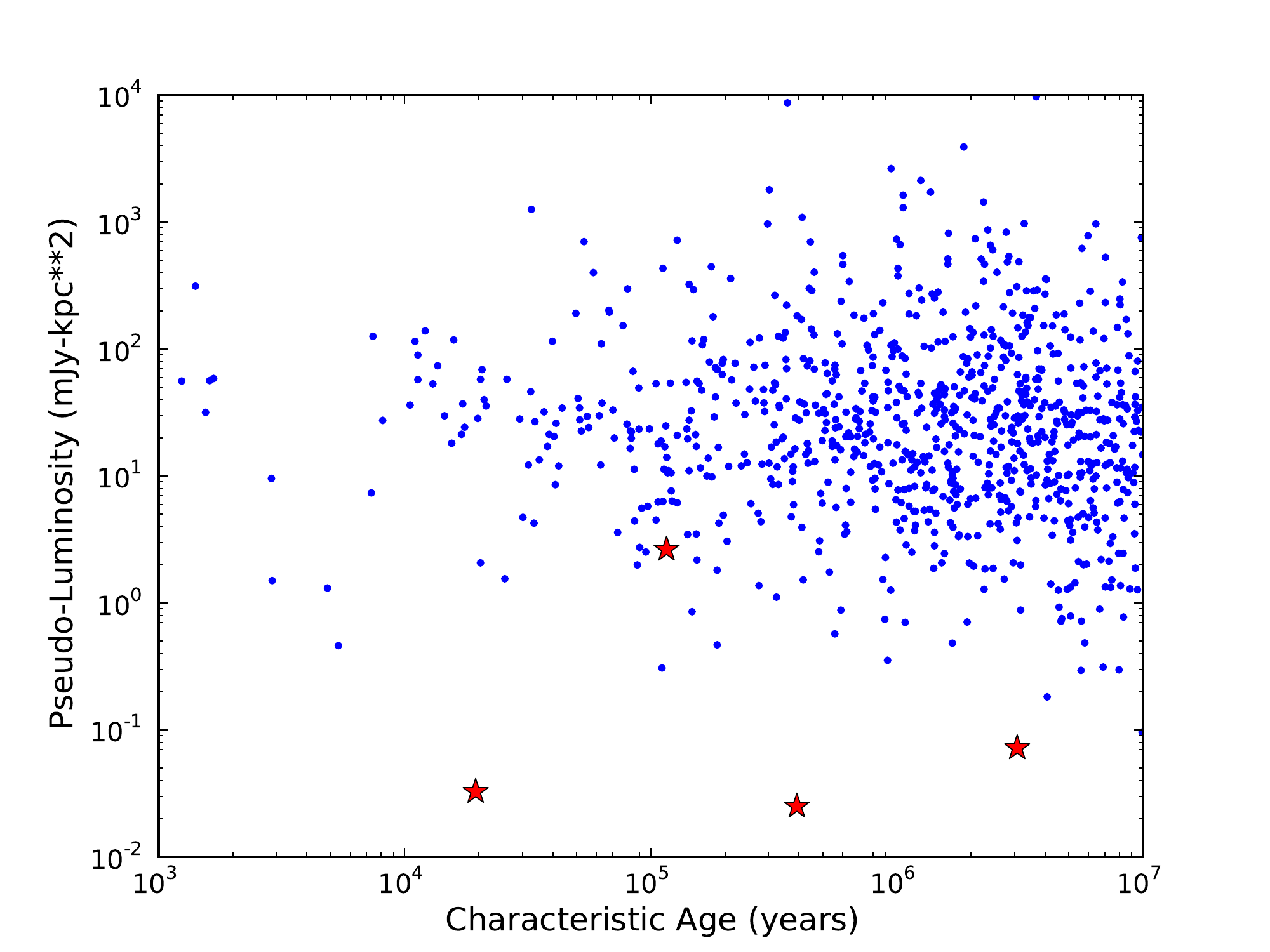}
\caption{Pseudo luminosities ($S_{1400} d^2$) for the four LAT blind search pulsars with detected radio pulsations, compared to the population of young pulsars in the ATNF catalog (blue dots). Three of them are less luminous than any previously known young pulsar.\label{fig:lum}}
\end{figure}

For the 4 radio detections we have DM distances and can thus 
convert the fluxes into pseudo-luminosities (in units of mJy-kpc$^2$) and these are shown in Figure \ref
{fig:lum}. Here we find that one of our pulsars (J2032+4127) is a modestly low luminosity radio pulsar 
while the other three have luminosities of a few $\times 10^{-2}$ mJy-kpc$^2$, far less luminous than any 
other normal pulsar. This forces us to consider what ``radio quiet'' really means and how pulsars could 
have such a low luminosity. Could we be just scraping the edge of the beam or is there some other effect at 
work?  To really answer these questions, we will need more examples of this low luminosity population and 
detailed modeling of the emission region incorporating geometric information from polarization 
measurements.

\section{Searching for Radio Pulsars in LAT Unassociated Sources}

\begin{figure}
\includegraphics[width=3.5in]{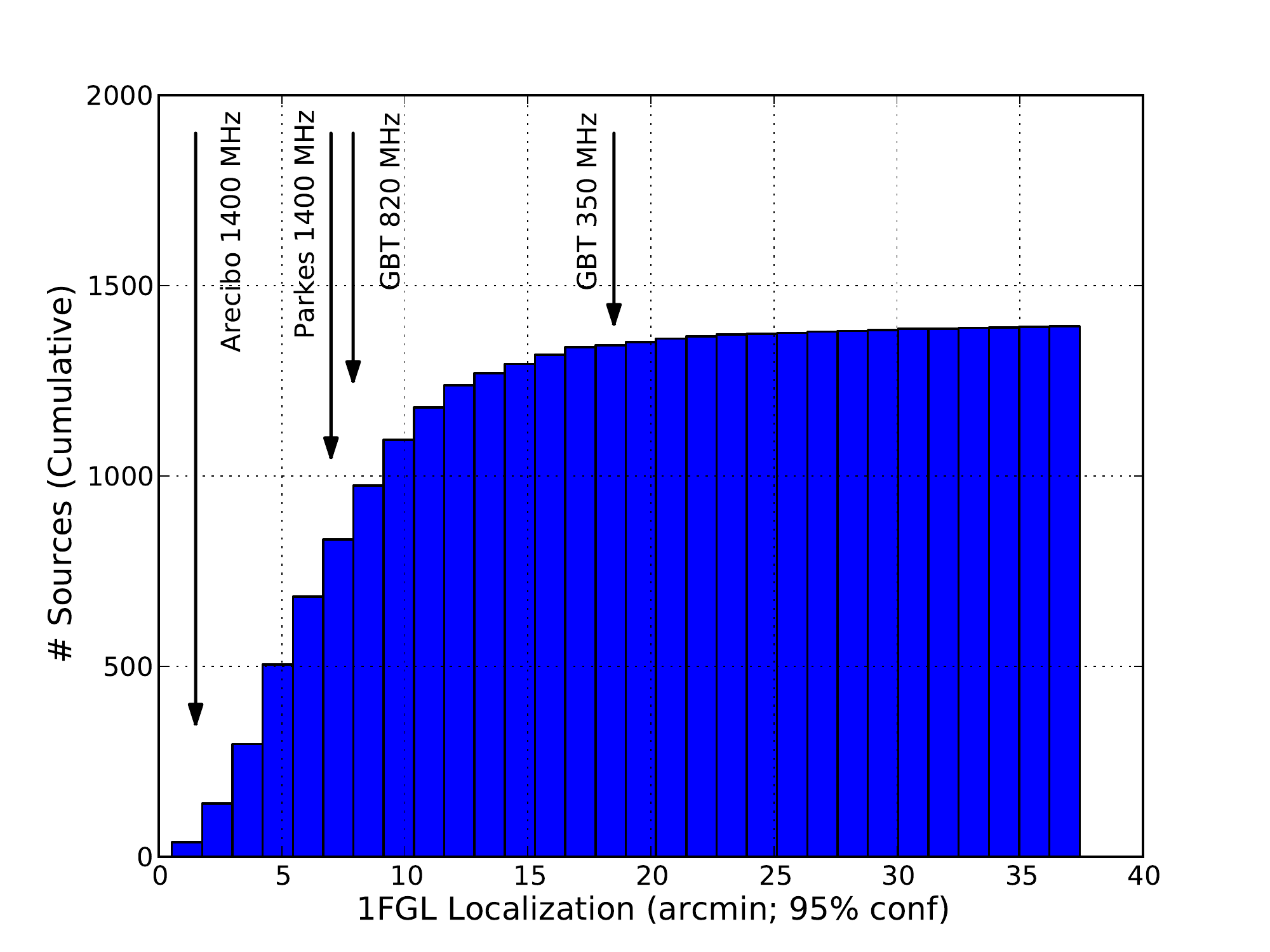}
\caption{Cumulative number of sources by 95\% confidence localization in the 1FGL catalog.  The FWHM beam widths of several telescope/frequency combinations are labeled.  Notice that Parkes at 1400 MHz and GBT at 820 MHz can cover 800--1000 of the 1400 sources in a single pointing.  By going to 350 MHz, even the 100-m GBT can cover the full error circle for nearly every source in the 1FGL catalog. The GMRT (the most recent addition to the PSC) has an extremely large beam width of $40'$ at 610 MHz and $80'$ at 325 MHz! \label{fig:loc}}
\end{figure}

With known pulsars being such a prevalent class of GeV $\gamma$-ray sources, a complementary strategy is to use radio telescopes to search for pulsars powering sources discovered in the LAT sky survey. A significant effort was put into searches like these during the EGRET era, with only modest success \cite{c06}. However, the sensitivity and angular resolution of the LAT make these searches much more powerful. The sensitivity provides many more sources -- there were $>600$ unassociated sources in the First LAT Source Catalog \cite{a+10b,a+12} -- and the angular resolution is such that most sources are localized to a region small enough to be covered by a single pointing of a large single-dish radio telescope (see Figure \ref{fig:loc}).

The PSC has undertaken a large number of pulsar searches directed at pulsar-like LAT sources.  Targets are ranked by how much they look like pulsars in their spectral and variability characteristics (see \cite{a+12} for the statistical methods used).  Initially, targets were selected at both high and low Galactic latitude, but it was quickly realized that the high latitude searches were highly successful at finding new millisecond pulsars, while the low latitude searches found only 1 new pulsar associated with a LAT source (PSR J2030+3641 \cite{ckr12}) and 3 slow pulsars that are almost certainly chance coincidences unrelated to the LAT sources (see Table \ref{tab:slow}).

The lack of success at finding young or middle-aged pulsars in searches of LAT sources in the plane is discussed in \cite{ckr12}. The basic argument is that most detectable normal radio pulsars in the plane have been discovered in large area surveys.  Thus, many of the pulsar-like $\gamma$-ray sources discovered by the LAT near the plane may indeed be pulsars, but with the radio beams not intersecting our line-of-sight. This is supported by the very small fraction (4/35) of blind search pulsars from which radio pulsations have been discovered.

\begin{figure*}
\includegraphics[width=170mm]{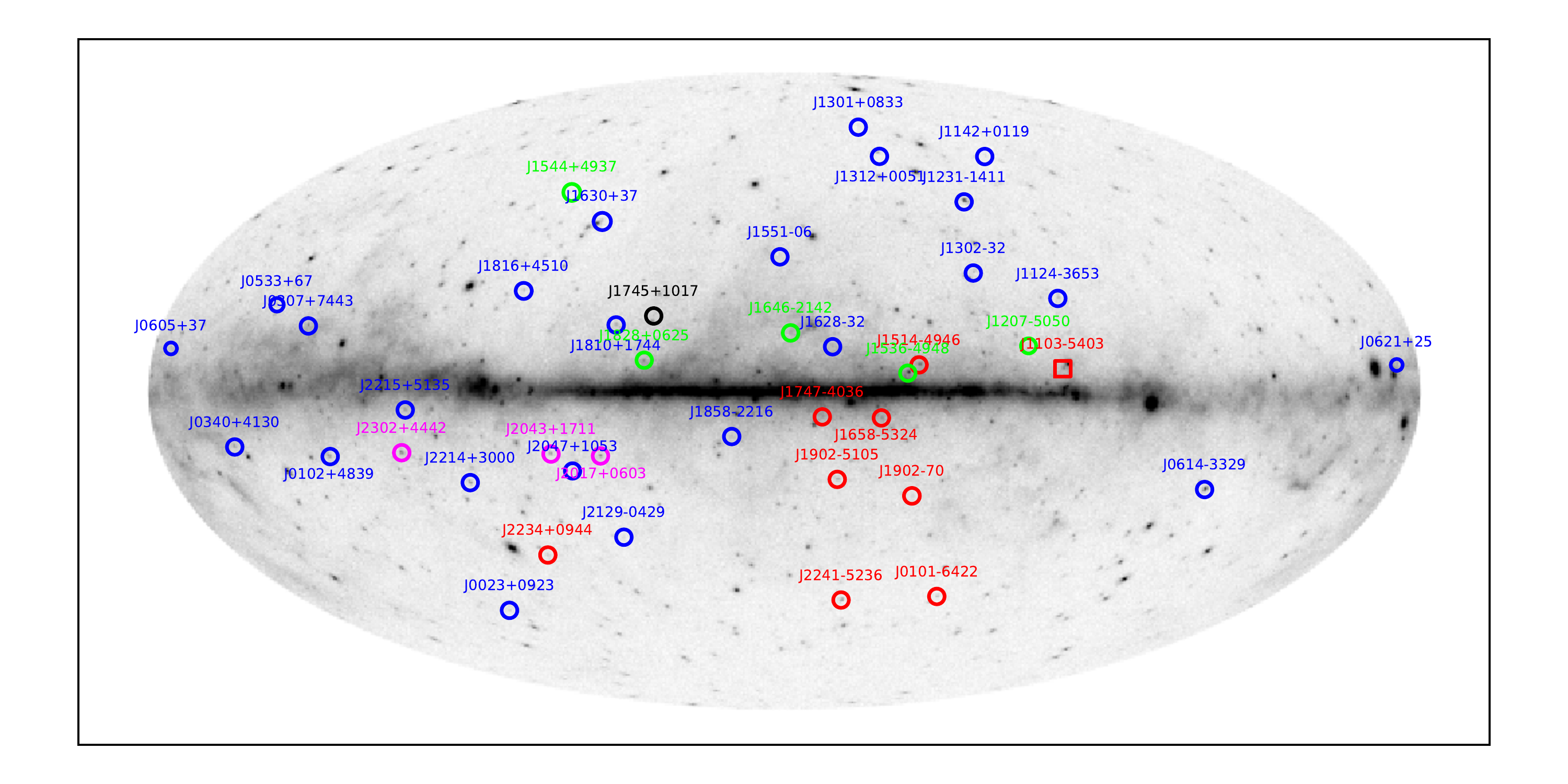}
\caption{\textit{Fermi} LAT sky map in Galactic coordinates showing the locations of the 43 MSPs discovered in searches of LAT unassociated source error circles. The symbols are color coded by the discovery telescope: blue for GBT, red for Parkes, green for GMRT, magenta for Nan\c{c}ay, black for Effelsberg. One GMRT discovery (J1124-36) is not shown since it overlaps in position with J1124-3653. The one square red symbol for J1103-5403 indicates that it was a serendipitous discovery of an MSP not associated with the LAT source \cite{kjr+11}. One of the GBT pulsars (J1816+4510) was discovered by the GBNCC team (PI: S. Ransom) in an analysis of survey beams coincident with LAT catalog sources \cite{s+12}.\label{fig:skymap}}
\end{figure*}

The situation at high latitude has been completely different.  At the end of the EGRET era, there were no confirmed $\gamma$-ray emitting MSPs.  Shortly after launch, the LAT demonstrated that MSPs were indeed $\gamma$-ray emitters, much like their young pulsar brethren \cite{a+09b}.  Our radio searches of high-latitude LAT sources that are non-variable and have curved spectra characteristic of pulsars have yielded 43 new MSPs (42 were discovered by the PSC and one by a non-PSC group also targeting LAT source; see Table \ref{tab:MSPs} and Figure \ref{fig:skymap}). 

Also, in contrast to the situation for young pulsars in the plane, the sky at high latitudes has mostly not been surveyed with good sensitivity to millisecond pulsars.  This is changing with several new radio surveys designed to detect MSPs (e.g. HTRU, GBT Drift, GBNCC).  In several cases, new MSPs discovered in these surveys have turned out to be LAT pulsars as well \cite{kjb+12}, and more are sure to come.

As the newly-discovered pulsars are timed over a year or so after discovery, we obtain orbital parameters and precise positions enabling us to fold the LAT $\gamma$-ray data to look for pulsations that confirm the association with the $\gamma$-ray source. In nearly all cases where phase-coherent timing solutions have been obtained the associations have been confirmed (as marked in Table \ref{tab:MSPs}), but in one case (PSR J1103$-$5403) the pulsar was not associated with the LAT source but just in the radio beam by chance. This is likely the case for more of the new discoveries with the GMRT, owing to its very large beam size.

One striking feature of this new population of MSPs discovered in $\gamma$-ray directed searches is the fraction in `black widow' (pulsars in orbit around very low mass companions and usually eclipsing) and `redback' (eclipsing binaries with a main sequence companion; see \cite{r11}) systems.  At the time of the \textit{Fermi} launch, there were about 60 MSPs known outside of the globular cluster system.  These included 3 black widows and 1 redback. Among the 43 new pulsars found in these searches, there are at least 10 black widows and 4 redbacks (additions to these numbers are possible because some of the pulsars have not yet had their orbital parameters determined). One possible explanation for this is that it is a selection effect, where the energetic MSPs powering $\gamma$-ray sources are more likely to be in these types of interacting binary systems, or possibly that some of the $\gamma$-rays are coming from the interaction region itself (intra-binary shock).

\begin{figure}
\includegraphics[width=3.5in]{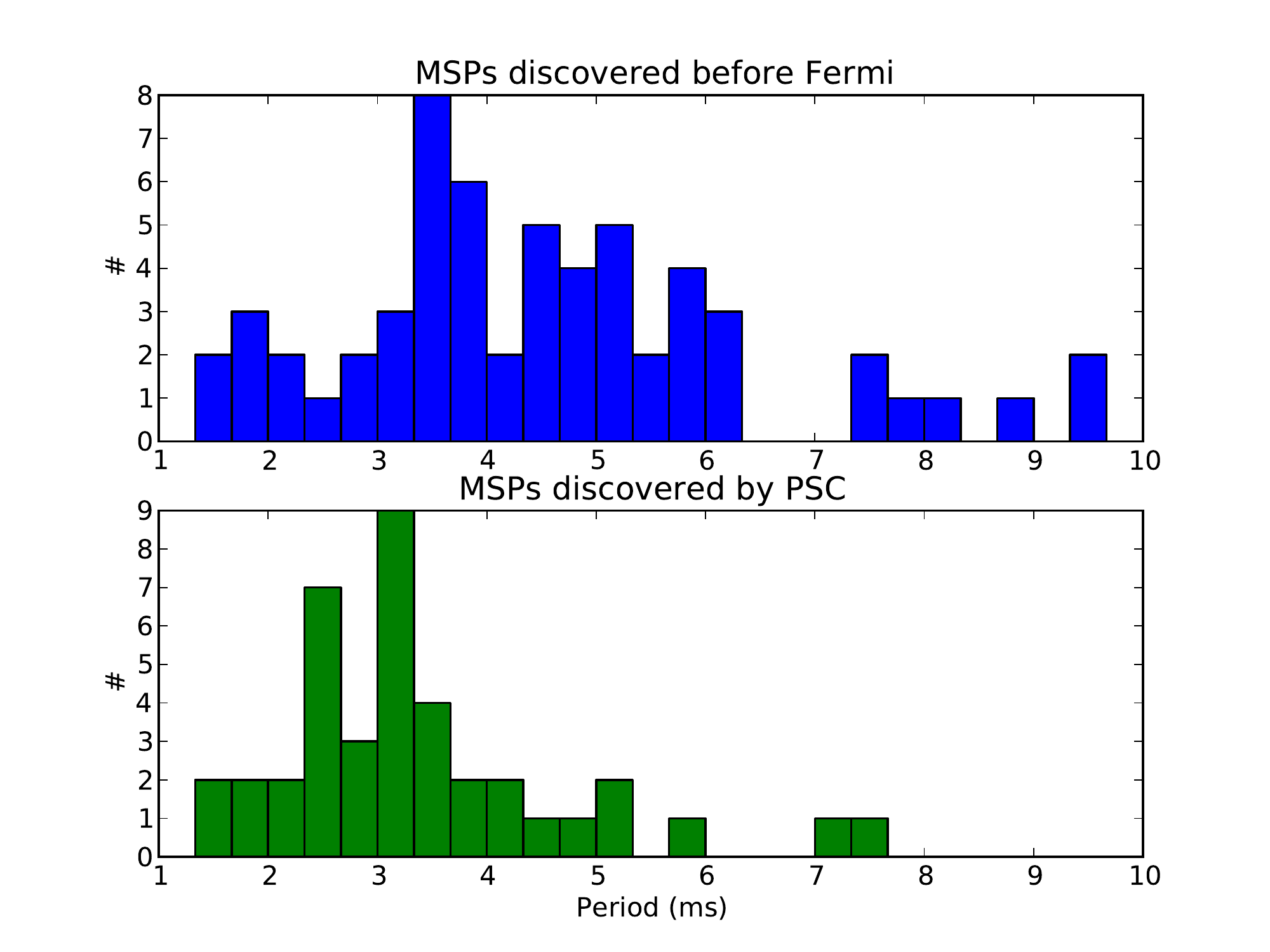}
\caption{Histograms of pulse periods of MSPs discovered before \textit{Fermi} and those discovered in PSC searches of LAT unassociated sources.  The selection for $\gamma$-ray emission selects for faster MSPs than large area radio surveys.\label{fig:hist}}
\end{figure}

Another indication that this population is somewhat different than the population discovered in large area radio surveys is the period distribution (see Figure~\ref{fig:hist}), which is much more strongly clustered at the short period end of the distribution than the general field MSP population (defined here as the 63 MSPs ($P<16$ ms) in the ATNF catalog with discovery dates before 2010 and not associated with globular clusters). The median periods of the two samples are 3.16 ms for the PSC MSPs and 4.62 ms for the previously known MSP population, and the K-S test give a probability of $1.3 \times 10^{-4}$ that they are drawn from the same underlying distribution. All of this supports earlier studies that find that the spin periods in eclipsing systems are preferentially faster than non-eclipsing systems \cite{h09}.

This large number of new MSPs also provides many potential new additions to pulsar timing array projects. Most of the eclipsing systems exhibit orbital period evolution and other effects that keep them from being among the best MSP clocks, but seven of the new systems are being timed by the NANOgrav\footnote{\url{http://nanograv.org}} project to evaluate their potential contribution to the detection of gravitational waves.

So, while these new MSPs are all radio pulsars that could \textit{eventually} have been discovered in undirected radio surveys, \textit{Fermi} has enabled us to discover them in a much more efficient manner and much earlier than they would have been, allowing for detailed follow-up observations and potentially speeding up the quest for the direct detection of gravitational waves. With different biases than the radio surveys, we are helping to get a more complete picture of the Galactic MSP population and finding many systems that will improve our understanding of MSP formation and evolution.  One lesson has been made very clear in our searches.  Each good candidate LAT source must be searched multiple times before declaring it to be devoid of a detectable radio pulsar.  In many cases the eclipses or strong interstellar scintillation prevent detection in a substantial fraction of the observations of a particular source.  Recently we have been focusing on making sure that the best candidate sources are observed 3, 4, or more times to account for this.

But, what about the pulsar-like high latitude sources where no radio pulsations are ever discovered?  These could be a population of radio quiet (as seen from Earth) MSPs.  Direct searches for millisecond pulsations in the LAT data are extremely computationally challenging, even for the $\sim$20\% of MSPs that are isolated \cite{g11,pga+12}.  For the binary MSPs, these searches are not currently tractable.  However, given the large fraction of black widow and redback systems among the LAT-detected MSPs, another approach is promising.  Optical and X-ray studies of the fields of the bright pulsar-like LAT source 0FGL J2339.8$-$0530 revealed a likely counterpart that is almost certainly an eclipsing binary MSP system \cite{rs11b,khc+12}.  With the orbital parameters and precise position determined from optical observations, a search for $\gamma$-ray pulsations from the pulsar in this system may be feasible.  X-ray and optical studies of the error boxes of other LAT sources may reveal more examples of this kind of system.

\begin{table}
\caption{Young Pulsars Discovered in PSC Searches\label{tab:slow}}
\begin{tabular}{lrcrr}
\hline
PSR & Period & DM  & LAT PSR? & Reference\\
    &   \multicolumn{1}{c}{(ms)}  & (pc cm$^{-3}$) & \\  
\hline\hline
J1604$-$44 & 1389.20 & 176 & No&\cite{kjr+11} \\
J2030+3641 &  200.13 & 247 & Yes &\cite{ckr12} \\
J1400$-$56 &  410.70 & 123 & No& \cite{k+12}\\
J1203$-$62 &  393.09 & 285 & No& \cite{k+12}\\
\hline
\end{tabular}
\end{table}

\begin{table*}
\caption{Millisecond pulsars discovered in radio pulsation searches of LAT unassociated sources \label{tab:MSPs}}
\begin{tabular}{lllrrrrrll}
\hline
PSR & Telescope & LAT Source & $P$  & DM  & $D$ & $P_\mathrm{orb}$ & $M_\mathrm{c,min}$  & Notes &Refs \\
 &  &  &  (ms) & (pc cm$^{-3}$) &  (kpc) &  (d) &  ($M_\odot$) &  \\
\hline\hline
J0023+0923 & GBT-350 & 1FGLJ0023.5+0930 & 3.05 & 14.3 & 0.7 & 0.14 & 0.017 & BW, $\gamma$ & \cite{hrm+11} \\
J0101$-$6422 & Parkes & 1FGLJ0101.0$-$6423 & 2.57 & 12.0 & 0.6 & 1.78 & 0.16 & $\gamma$ & \cite{kcj12}\\
J0102+4839 & GBT-350 & 1FGLJ0103.1+4840 & 2.96 & 53.5 & 2.3 & 1.67 & 0.18 & $\gamma$ &  \cite{hrm+11}\\
J0307+7443 & GBT-350 & 1FGLJ0308.6+7442 & 3.16 & 6.4 & 0.6 & 36.98 & 0.24 & $\gamma$ &\cite{hrm+11}\\
J0340+4130 & GBT-350 & 1FGLJ0340.4+4130 & 3.30 & 49.6 & 1.8 & Isolated &  & $\gamma$ &  \cite{hrm+11} \\
J0533+67 & GBT-820 & 2FGLJ0533.9+6759 & 4.39 & 57.4 & 2.4 & Isolated & \\
J0605+37 & GBT-820 & 2FGLJ0605.3+3758 & 2.73 & 21.0 & 0.7 & 55.6 & 0.18\\
J0614$-$3329 & GBT-820 & 1FGLJ0614.1$-$3328 & 3.15 & 37.0 & 1.9 & 53.6 & 0.28 & $\gamma$& \cite{rrc+11}\\
J0621+25 & GBT-820 & 2FGLJ0621.2+2508 & 2.72 & 83.6 & 2.3 & TBD & \\
J1103$-$5403 & Parkes & 1FGLJ1103.9$-$5355 & 3.39 & 103.9 & 2.5 & Isolated & & NA &  \cite{kjr+11}\\
J1124$-$3653 & GBT-350 & 1FGLJ1124.4$-$3654 & 2.41 & 44.9 & 1.7 & 0.23 & 0.027 & BW, $\gamma$ &  \cite{hrm+11}  \\
J1124$-$36  & GMRT-325 & 1FGLJ1124.4$-$3654 & 5.55 & 45.1 & 1.7 & TBD & & NA & \\
J1142+0119  & GBT-820 & 1FGLJ1142.7+0127 & 5.07 & 19.2 & 0.9 & 1.58 & 0.15\\
J1207$-$5050 & GMRT-610 & 1FGLJ1207.0$-$5055 & 4.84 & 50.6 & 1.5 & Isolated? & \\
J1231$-$1411 & GBT-820 & 1FGLJ1231.1$-$1410 & 3.68 & 8.1 & 0.4 & 1.86 & 0.19 & $\gamma$& \cite{rrc+11}\\
J1301+0833 & GBT-820 & 1FGLJ1301.8+0837 & 1.84 & 13.2 & 0.7 & 0.27 & 0.024 & BW &\\
J1302$-$32 & GBT-350 & 1FGLJ1302.3$-$3255 & 3.77 & 26.2 & 1.0 & 0.78 & 0.15 & &  \cite{hrm+11} \\
J1312+0051 & GBT-820 & 1FGLJ1312.6+0048 & 4.23 & 15.3 & 0.8 & 38.5 & 0.18 & $\gamma$\\
J1514$-$4946 & Parkes & 1FGLJ1514.1$-$4945 & 3.59 & 30.9 & 0.9 & 1.92 & 0.17 & $\gamma$  & \cite{kcj12}\\
J1536$-$4948 & GMRT-325 & 1FGLJ1536.5$-$4949 & 3.08 & 38.0 & 1.8 & TBD & \\
J1544+4937 & GMRT-610 & 18M3037 & 2.16 & 23.2 & 1.2 & 0.117 & 0.018 & BW & \cite{br+12} \\
J1551$-$06\S & GBT-350 & 1FGLJ1549.7$-$0659 & 7.09 & 21.6 & 1.0 & 5.21 & 0.20 & & \cite{hrm+11} \\
J1628$-$3205 & GBT-820 & 1FGLJ1627.8$-$3204 & 3.21 & 42.1 & 1.2 & 0.21 & 0.16 & RB &\\
J1630+37 & GBT-820 & 2FGLJ1630.3+3732 & 3.32 & 14.1 & 0.9 & 12.5 & 0.16\\
J1646$-$2142 & GMRT-325 & 1FGLJ1645.0$-$2155c & 5.85 & 29.8 & 1.1 & 23 & TBD \\
J1658$-$5324 & Parkes & 1FGLJ1658.8$-$5317 & 2.43 & 30.8 & 0.9 & Isolated & & $\gamma$ & \cite{kcj12} \\
J1745+1017 & Effelsberg & 1FGLJ1745.5+1018 & 2.65 & 23.9 & 1.3 & 0.73 & 0.014 & BW, $\gamma$ & \cite{b+12}\\
J1747$-$4036 & Parkes & 1FGLJ1747.4$-$4036 & 1.64 & 153.0 & 3.4 & Isolated &  & $\gamma$ & \cite{kcj12} \\
J1810+1744 & GBT-350 & 1FGLJ1810.3+1741 & 1.66 & 39.6 & 2.0 & 0.15 & 0.045 & BW, $\gamma$ &  \cite{hrm+11} \\
J1816+4510 & GBT-350 & 1FGLJ1816.7+4509 & 3.19 & 38.9 & 2.4 & 0.36 & 0.16 & $\dag$, RB, $\gamma$ & \cite{ks+12}  \\
J1828+0625 & GMRT-325 & 1FGLJ1830.1+0618 & 3.63 & 22.4 & 1.2 & 6.0 & TBD\\
J1858$-$2216 & GBT-820 & 1FGLJ1858.1$-$2218 & 2.38 & 26.6 & 0.9 & 46.1 & 0.22 & $\gamma$\\
J1902$-$5105 & Parkes & 1FGLJ1902.0$-$5110 & 1.74 & 36.3 & 1.2 & 2.01 & 0.16 & $\gamma$ & \cite{kcj12}\\
J1902$-$70  & Parkes & 2FGLJ1902.7$-$7053 & 3.60 & 19.5 & 0.8 & TBD & & \\
J2017+0603 & Nan\c{c}ay & 1FGLJ2017.3+0603 & 2.90 & 23.9 & 1.6 & 2.2 & 0.18 &$\gamma$ & \cite{cgj+11}\\
J2043+1711 & Nan\c{c}ay & 1FGLJ2043.2+1709 & 2.38 & 20.7 & 1.8 & 1.48 & 0.17 & $\gamma$ & \cite{gfc+12}\\
J2047+1053 & GBT-820 & 1FGLJ2047.6+1055 & 4.29 & 34.6 & 2.0 & 0.12 & 0.036 & BW, $\gamma$ & \\
J2129$-$0429 & GBT-350 & 1FGLJ2129.8$-$0427 & 7.62 & 16.9 & 0.9 & 0.64 & 0.37 & RB & \cite{hrm+11}  \\
J2214+3000 & GBT-820 & 1FGLJ2214.8+3002 & 3.12 & 22.5 & 1.5 & 0.42 & 0.014 & BW, $\gamma$ & \cite{rrc+11}\\
J2215+5135 & GBT-350 & 1FGLJ2216.1+5139 & 2.61 & 69.2 & 3.0 & 0.17 & 0.21 & RB, $\gamma$ &  \cite{hrm+11} \\
J2234+0944 & Parkes & 1FGLJ2234.8+0944 & 3.63 & 17.8 & 1.0 & 0.42 & 0.015 & BW, $\gamma$ & \cite{k+12} \\
J2241$-$5236 & Parkes & 1FGLJ2241.9$-$5236 & 2.19 & 11.5 & 0.5 & 0.14 & 0.012 & BW, $\gamma$ & \cite{kjr+11} \\
J2302+4442 & Nan\c{c}ay & 1FGLJ2302.8+4443 & 5.19 & 13.8 & 1.2 & 125.9 & 0.3 & $\gamma$& \cite{cgj+11}\\
\hline
\end{tabular}
\\
Notes: The number after the telescope name indicates the observing frequency in MHz; no number means 1400 MHz. \S = This pulsar was previously published as J1549$-$06 \cite{hrm+11,n+12} before the precise position was determined. \\ BW = black widow, RB = redback, NA = known to not be associated with the target LAT source, \\ $\dag$ = Discovered by GBNCC team in analysis of beams coincident with LAT sources \cite{s+12}, $\gamma$ = $\gamma$-ray pulsations detected
\end{table*}

\bigskip 
\begin{acknowledgments}

The \textit{Fermi} LAT Collaboration acknowledges support from a number of agencies and 
institutes for both development and the operation of the LAT as well as scientific data 
analysis. These include NASA and DOE in the United States, CEA/Irfu and IN2P3/CNRS in 
France, ASI and INFN in Italy, MEXT, KEK, and JAXA in Japan, and the K.~A.~Wallenberg 
Foundation, the Swedish Research Council and the National Space Board in Sweden. 
Additional support from INAF in Italy and CNES in France for science analysis during the 
operations phase is also gratefully acknowledged.

The Parkes radio telescope is part of the Australia Telescope which is funded by the Commonwealth Government for operation as a National Facility managed by CSIRO. 

The Green Bank Telescope is operated by the National Radio Astronomy Observatory, a facility of the National Science Foundation operated under cooperative agreement by Associated Universities, Inc.

The Arecibo Observatory is a facility of the NSF operated by SRI International, Universities Space Research Association and Universidad Metropolitana.

The Nan\c{c}ay Radio Observatory is operated by the Paris Observatory, associated with the French Centre National de la Recherche Scientifique (CNRS).

The GMRT is run by the National Centre for Radio Astrophysics (NCRA) of the 
Tata Institute of Fundamental Research (TIFR).

\end{acknowledgments}

\bigskip 


\begin{thebibliography}{9} 

\bibitem{a+09}
Abdo, A. A., et al. 2009, ``Detection 
of 16 Gamma-Ray Pulsars Through Blind Frequency Searches Using the Fermi 
LAT'', Science,  325,  840 

\bibitem{a+09b}
Abdo, A. A., et al. 2009,  ``A 
Population of Gamma-Ray Millisecond Pulsars Seen with the Fermi Large Area 
Telescope'', Science,  325,  848

\bibitem{a+10}
Abdo, A. A., et al. 2010,  
``PSR J1907+0602: A Radio-Faint Gamma-Ray Pulsar Powering a Bright TeV 
Pulsar Wind Nebula'', ApJ,  711,  64 

\bibitem{a+10b}
Abdo, A.~A., et al.\ 2010, ``Fermi Large Area Telescope First Source Catalog'', ApJS, 188, 405 

\bibitem{a+12}
Abdo, A. A. et al. 2012, ``A Statistical Approach to Recognizing Source Classes for Unassociated Sources in the First Fermi-LAT Catalog'', ApJ, in press (arXiv:1108.1202)

\bibitem{b+12}
Barr, E. D. et al., MNRAS 2012, in prep.

\bibitem{br+12}
Bhattacharyya et al., 2012 in prep.

\bibitem{crr+09}
Camilo, F., Ray, P. S., Ransom, et al. 2009,  ``Radio 
Detection of LAT PSRs J1741-2054 and J2032+4127: No Longer Just Gamma-ray 
Pulsars'', ApJ,  705,  1 

\bibitem{ckr12}
Camilo, F., Kerr, M., Ray, P. S., et al. 2012, ``PSR J2030+3641: 
Radio Discovery and Gamma-Ray Study of a Middle-aged Pulsar in the Now 
Identified Fermi-LAT Source 1FGL J2030.0+3641'', ApJ, 746,  39

\bibitem{cgj+11}
Cognard, I., Guillemot, L., Johnson, T. J., et al. 2011,  
``Discovery of Two Millisecond Pulsars in Fermi Sources with the Nan\c{c}ay 
Radio Telescope'', ApJ,  732,  47 

\bibitem{c06}
Crawford, F. et al. 2006, ApJ, 652, 1499

\bibitem{g11} 
Guillemot, L. 2011, ``A blind search for isolated millisecond pulsars in the Fermi LAT data'', poster PSR S2.N8 at Fermi Symposium 2011, Rome, Italy

\bibitem{gfc+12}
Guillemot, L., Freire, P. C. C., Cognard, I. et al. 2012, ``Discovery of the millisecond pulsar PSR J2043+1711 in a Fermi source with the Nan\c{c}ay Radio Telescope'', MNRAS, 422, 1294

\bibitem{h09}
Hessels, J. W. T. 2009, ``The Observed Spin Distributions of Millisecond Radio and X-ray Pulsars'', AIP Conference Proc. (arXiv:0903.0493)

\bibitem{mad12}
Maan, Y., Aswathappa, H. A. \& Deshpande, A. A. 2012, ``Pulsed Radio Emission from the Fermi-LAT Pulsar J1732-3131: Search and A Possible Detection at 34.5 MHz'', MNRAS, in press (arXiv:1109.6032)

\bibitem{hrm+11}
Hessels, J. W. T., Roberts, M. S. E., McLaughlin, M. A., Ray, P. S., 
Bangale, P., Ransom, S. M., Kerr, M., Camilo, F., DeCesar, M. E. 2011,  
``A 350-MHz GBT Survey of 50 Faint Fermi $\gamma$-ray Sources for Radio Millisecond 
Pulsars'', AIP Conference Series,  1357,  40

\bibitem{ks+12}
Kaplan, D. et al. 2012, ApJ, submitted

\bibitem{kjr+11}
Keith, M. J., Johnston, S., Ray, P. S., et al. 2011,  
``Discovery of millisecond pulsars in radio searches of southern Fermi Large 
Area Telescope sources'', MNRAS, 414,  1292 

\bibitem{kjb+12}
Keith, M. J., et al. 2012,  ``The 
High Time Resolution Universe Pulsar Survey - IV. Discovery and polarimetry 
of millisecond pulsars'', Monthly Notices of the Royal Astronomical Society,  
419, 1752

\bibitem{k+12}
Keith, M. J. et al. 2012, in preparation

\bibitem{kcj12}
Kerr, M., Camilo, F., Johnson, T. J., et al. 2012, ``Five 
New Millisecond Pulsars From a Radio Survey of 14 Unidentified Fermi-LAT 
Gamma-ray Sources'', ApJ, 748, L2

\bibitem{khc+12}
Kong, A. K. H., et al. 2012,  
``Discovery of an Unidentified Fermi Object as a Black Widow-like 
Millisecond Pulsar'', ApJ,  747,  L3 

\bibitem{n+12} Nolan, P.~L., et al.\ 2012, ApJS, 199, 31 

\bibitem{pga+12}
Pletsch, H. J., Guillemot, L., Allen, B., et al. 2012,  ``Discovery 
of Nine Gamma-Ray Pulsars in Fermi Large Area Telescope Data Using a New 
Blind Search Method'', ApJ,  744,  105 

\bibitem{rc69}
Radhakrishnan, V. and Cooke, D. J. 1969,  ``Magnetic Poles and the 
Polarization Structure of Pulsar Radiation'', Astrophysical Letters,  3,  
225 

\bibitem{rrc+11}
Ransom, S. M., Ray, P. S., Camilo, F., et al. 2011,  ``Three Millisecond 
Pulsars in Fermi LAT Unassociated Bright Sources'', ApJ,  727,  L16 

\bibitem{rkp+11}
Ray, P. S., Kerr, M., Parent, D., et al. 2011,  ``Precise $\gamma$-ray Timing 
and Radio Observations of 17 Fermi $\gamma$-ray Pulsars", ApJS,  194,  17 

\bibitem{rs11}
Ray, P. S., Saz Parkinson, P. M. 2011,  ``Pulsar Results with the 
Fermi Large Area Telescope'', in \textit{High-Energy Emission from Pulsars and their 
Systems}, Astrophysics and Space Science Proceedings, 37

\bibitem{r11}
Roberts, M. S. E. 2011, ``New Black Widows and Redbacks in the Galactic Field'', in AIP Conf. Ser. 1357, 127 (arXiv:1103.0819)

\bibitem{rs11b}
Romani, R. W., Shaw, M. S. 2011,  ``The Orbit and Companion of 
Probable $\gamma$-Ray Pulsar J2339$-$0533'', ApJ,  743,  L26

\bibitem{rw10}
Romani, R. W., Watters, K. P. 2010,  ``Constraining Pulsar 
Magnetosphere Geometry with $\gamma$-ray Light Curves'', ApJ,  
714,  810 

\bibitem{sdz+10}
Saz Parkinson, P. M., Dormody, M., Ziegler, M., Ray, P. S., et al. 2010,  ``Eight 
$\gamma$-ray Pulsars Discovered in Blind Frequency Searches of Fermi LAT Data'', 
ApJ,  725,  571 

\bibitem{sp+12}
Saz Parkinson, P. M. et al. 2012, ``Two Additional Pulsars Discovered in Blind Frequency Searches'', in prep.

\bibitem{s+12}
Stovall, K. 2012, in prep.

\bibitem{twc11}
Takata, J., Wang, Y., and Cheng, K. S. 2011,  ``Population study for $\gamma$-ray 
pulsars with the outer-gap model - III. Radiation characteristics and 
viewing geometry'', MNRAS,  415,  1827 

\bibitem{wr11}
Watters, K. P., Romani, R. W. 2011,  ``The Galactic Population of 
Young $\gamma$-ray Pulsars'', ApJ,  727,  123 

\end{thebibliography}
\end{document}